\documentclass{article}

    \PassOptionsToPackage{numbers, compress}{natbib}

\usepackage[final]{neurips_2025}

\usepackage[utf8]{inputenc} 
\usepackage[T1]{fontenc}    
\usepackage[hidelinks]{hyperref}       
\usepackage{url}            
\usepackage{booktabs}       
\usepackage{amsfonts}       
\usepackage{amsmath}        
\usepackage{nicefrac}       
\usepackage{microtype}      
\usepackage{xcolor}         
\usepackage{graphicx}
\usepackage{subcaption}
\usepackage{caption}
\usepackage{float}
\usepackage{tcolorbox}
\usepackage{fancyvrb}

\usepackage[dvipsnames]{xcolor}

\graphicspath{{figures/}}
\title{\textsc{SketchMind}:  A Multi-Agent Cognitive Framework for Assessing Student-Drawn Scientific Sketches}

\author{%
    Ehsan Latif \\
    AI4STEM Education Center\\
    University of Georgia\\
    Athens, GA 30605
    \And
    Zirak Khan \\
    School of Computing\\
    University of Georgia\\
    Athens, GA 30605
    \And
    Xiaoming Zhai \thanks{Corresponding author email: \texttt{xiaomig.zhai@uga.edu}} \\
    AI4STEM Education Center\\
    University of Georgia\\
    Athens, GA 30605
}

\begin{document}

\maketitle

\begin{abstract}
  Scientific sketches (e.g., models) offer a powerful lens into students' conceptual understanding, yet AI-powered automated assessment of such free-form, visually diverse artifacts remains a critical challenge. Existing solutions often treat sketch evaluation as either an image classification task or monolithic vision-language models, which lack interpretability, pedagogical alignment, and adaptability across cognitive levels. To address these limitations, we present \textsc{SketchMind}, a cognitively grounded, multi-agent framework for evaluating and improving student-drawn scientific sketches. \textsc{SketchMind} introduces \emph{Sketch Reasoning Graphs (\texttt{SRGs)}}, semantic graph representations that embed domain concepts and Bloom's taxonomy-based cognitive labels. The system comprises modular agents responsible for rubric parsing, sketch perception, cognitive alignment, and iterative feedback with sketch modification, enabling personalized and transparent evaluation. We evaluate \textsc{SketchMind} on a curated dataset of 3,575 student-generated sketches across six science assessment items with different highest order of Bloom's level that require students to draw models to explain phenomena. Compared to baseline GPT-4o performance without \texttt{SRG} (average accuracy: 55.6\%), and with \texttt{SRG} integration achieves 77.1\% average accuracy (+21.4\% average absolute gain). We also demonstrate that multi-agent orchestration with \texttt{SRG} enhances \textsc{SketchMind} performance, for example, GPT-4.1 gains an average 8.9\% increase in sketch prediction accuracy, outperforming single-agent pipelines across all items. Human evaluators rated the feedback and co-created sketches generated by \textsc{SketchMind} with GPT-4.1, which achieved an average of 4.1 out of 5, significantly higher than those of baseline models (e.g., 2.3 for GPT-4o). Experts noted the system’s potential to meaningfully support conceptual growth through guided revision. Our code and (pending approval) dataset will be released to support reproducibility and future research in AI-driven education.

\end{abstract}

\section{Introduction}

Sketching such as drawn models is a fundamental tool in science education, allowing students to externalize their thinking, represent causal mechanisms, and engage in higher-order reasoning \cite{zhai2022applying}. However, assessing the quality and cognitive depth of student-generated sketches remains a long-standing challenge due to their open-ended nature and semantic variability. Automated systems often struggle with interpreting free-form, domain-rich visual input, which makes effective feedback and evaluation particularly difficult. Recent advances in multimodal large language models (MLLMs), such as GPT-4V, have enabled breakthroughs in vision-language reasoning, leading to promising applications in educational assessment \cite{lee2023gemini}. Specifically, systems like NeRiF have shown that MLLMs can approximate expert grading of student-drawn models by extracting latent structure from images and comparing them against rubrics \cite{lee2023nerif}. However, such monolithic models still face limitations regarding their reasoning processes, feedback personalization, and inconsistent across conceptually diverse tasks \cite{yaacoub2025assessing, hong2024my}.

To address these limitations, we propose \textsc{SketchMind}, a cognitively grounded, multi-agent framework for evaluating and enhancing scientific sketches. \textsc{SketchMind} models sketches as \texttt{SRG}s, which embed both structural semantics and Bloom's Taxonomy-based cognitive annotations \cite{hui2025incorporating, gonsalves2024generative, elim2024promoting}. Each \texttt{SRG} encodes domain concepts, their relationships, and Bloom's levels, enabling meaningful alignment with rubrics and scaffolding-targeted feedback. Inspired by the pedagogical principles behind systems like Betty’s Brain \cite{biswas2005learning, leelawong2008designing, biswas2016design}, \textsc{SketchMind} framework decomposes the assessment task across four specialized agents. These agents perform (1) rubric parsing and \texttt{SRG} generation, (2) sketch perception and \texttt{SRG} inference, (3) cognitive alignment and scoring, and (4) iterative feedback and sketch modification. This modular design is grounded in cognitive science and agentic learning principles, enabling transparent reasoning and pedagogically informed intervention \cite{kamalov2025evolution, guo2024using}.


Through extensive evaluation on a curated NGSS-aligned dataset of student-drawn science sketches \cite{zhai2022applying}, we show that \textsc{SketchMind} not only improves the baseline monolithic MLLM approaches but also increases the capabilities of reasoning models in both quantitative metrics (accuracy and alignment) and qualitative human feedback (clarity, relevance, pedagogical value). Human experts highlighted that \textsc{SketchMind} with models like GPT-4.1 can iteratively improve students' conceptual understanding and sketch quality via visual hints and structured revision cycles. Here are key contributions of this paper:
\begin{itemize}
    \item We introduce \textsc{SketchMind}, a multi-agent framework that integrates cognitive theories with AI to assess student-generated sketches effectively using our proposed \texttt{SRG}s.
    \item We integrate Bloom’s Taxonomy as cognition theory standard to construct and analyze \texttt{SRG}s for structured evaluation of visual student work and provides pedagogically sound feedback along with real-time sketch modification.
    \item With empirical studies, we found that \textsc{SketchMind} with State-Of-The-Art MLLM such as GPT-4.1, is able to achieve an average 90.2\% sketch prediction accuracy and generate pedagologically sound feedback with sketch modifications highly rated by human experts (4.1 out of 5), highlighting its potential for advanced AI-supported learning of scientific concepts.
\end{itemize}

To promote transparency and facilitate further research, we have open-sourced our codebase at Anonymzed repository\footnote{\url{https://anonymous.4open.science/r/SRG-287B}} and plan to make the dataset publicly available upon receiving the necessary approvals. This work represents a step forward in AI for Education, demonstrating how cognitively-aware, agentic systems can advance the quality, transparency, and effectiveness of automated reasoning over student-generated visual content.

\section{Related Work}

\textbf{Sketch Understanding and Visual Reasoning.}
Recent years have seen significant progress in sketch understanding, particularly within the computer vision community. Approaches such as SketchFusion \cite{bhunia2023sketch2saliency}, Sketch2Saliency \cite{bhunia2023sketch2saliency}, and SketchXAI \cite{qu2023sketchxai} have explored the utility of human-drawn sketches for learning visual concepts and providing interpretable representations. These works primarily focus on object detection, image retrieval \cite{chaudhuri2023data}, or 3D modeling \cite{mikaeili2023sked}, rather than assessing the conceptual depth embedded in scientific sketches. Educationally focused sketch models such as SEVA \cite{mukherjee2023seva} and DrawEduMath \cite{baral2025drawedumath} analyze human abstraction or math reasoning but lack cognitive scaffolding like Bloom’s taxonomy. \textsc{SketchMind} departs from these efforts by representing student-drawn sketches as cognitively annotated semantic graphs and grounding visual elements in educational rubrics, allowing for pedagogical interpretation and targeted feedback.

\textbf{Multimodal and Agentic Reasoning in Education.}
MLLMs such as GPT-4V have opened new opportunities in visual question answering and diagrammatic reasoning \cite{lee2023gemini}. While tools like NeRiF \cite{lee2023nerif} demonstrate GPT-4V's ability to grade drawn models, these systems often operate as monolithic black boxes, limiting transparency and pedagogical adaptability. Similarly, recent work in multimodal chain-of-thought reasoning \cite{zhai2022applying, xu2025towards, yin2025clearsight} and multi-agent systems \cite{geng2025realm, guo2024using, shahzad2024multi, kamalov2025evolution} show promise for decomposing complex tasks. However, most frameworks either lack cognitive modeling or fail to integrate sketch modifications explicitly. In contrast, \textsc{SketchMind} brings together modular reasoning agents with fine-grained cognitive alignment, enabling both transparent evaluation and actionable feedback.

\textbf{Educational AI for Scientific Sketch Assessment.}
Several studies have explored automated grading and classification of hand-drawn sketches in educational settings. Rakhmanov \cite{Rakhmanov2020ANAA} proposed a quality-based classification framework for freehand sketches, while Rahaman et al. \cite{Rahaman2024AutomatedGAA} applied CNN-based models for accuracy prediction. Lee et al. \cite{Lee2023AutomatedAOA} developed a rubric-driven grading system focused on particulate matter diagrams. These systems, however, primarily emphasize surface-level visual features and often lack semantic or cognitive interpretation. More recent work has tried to integrate learning objectives, such as \cite{yaacoub2025assessing, hong2024my}, yet they remain constrained to textual responses or fixed rubrics. \textsc{SketchMind} bridges this gap by introducing \texttt{SRG}s embedded with Bloom-level annotations and enabling multi-agent-driven sketch modification.

\section{Proposed Approach}
\label{sec:proposed_approach}

We present \textsc{SketchMind}, a cognitively grounded, multi-agent framework for evaluating and improving scientific sketches through iterative, feedback-driven modification. \textsc{SketchMind} is anchored in \emph{Bloom’s Taxonomy}~\cite{hui2025incorporating}, a hierarchical model of cognitive processes ranging from recall to creative synthesis. By modeling sketches as semantic structures called \texttt{SRG}s, \textsc{SketchMind} align student-generated content with domain rubrics and provide interpretable, formative feedback across cognitive levels.

\subsection{Cognitive Framework: Bloom’s Taxonomy in Sketch Understanding}

Bloom’s Taxonomy structures learning objectives into six ascending levels of cognitive complexity \cite{elim2024promoting}:
\begin{equation}
\mathcal{B} = \{\mathrm{Remember}, \mathrm{Understand}, \mathrm{Apply}, \mathrm{Analyze}, \mathrm{Evaluate}, \mathrm{Create}\}.
\end{equation}
These range from basic recall of knowledge (\textsc{Remember}) to the synthesis of novel ideas (\textsc{Create}). This cognitive hierarchy has long served as a foundation in science education for designing assessments and scaffolding learning \cite{hui2025incorporating}. Prior work further shows that aligning instructional technologies with Bloom’s levels supports measurable gains in higher-order thinking \cite{gonsalves2024generative}. In \textsc{SketchMind}, scientific sketches are not merely visual representations but are conceptualized as cognitive artifacts that externalize learners’ mental models. To operationalize this, we annotate each node (concept) and edge (relation) in the Sketch Response Graph (\texttt{SRG}) with a Bloom level, a process we call \textit{Bloom-Level Annotation}. This provides a fine-grained measure of the depth of conceptual engagement demonstrated by the student.


To implement this systematically, rubric statements are parsed to extract key verbs and criteria, which are then mapped to Bloom levels using a curated lexicon \cite{reddy2010review}. Ambiguous cases are resolved with a fine-tuned classifier. The resulting numeric levels (1 for \textsc{Remember} through 6 for \textsc{Create}) are encoded as attributes on SRG nodes and edges. These attributes directly support semantic similarity scoring, nuanced evaluation, and feedback generation.

\paragraph{Pedagogical Integration.}  
Each gold-standard SRG is labeled at the level of its highest Bloom demand. Student sketches are then evaluated by comparing the Bloom-level annotations of their SRGs against this reference. This supports both diagnostics (identifying the highest level achieved) and adaptive feedback. For example, if a student’s sketch demonstrates \textsc{Understand}, the system can generate targeted prompts or visual hints nudging them toward \textsc{Apply}. Such progression-oriented scaffolding is consistent with educational research showing that adaptive support aligned with Bloom’s hierarchy fosters deeper learning in STEM domains \cite{hui2025incorporating, gonsalves2024generative}.

A schematic illustration (see Fig.~\ref{fig:SRG_overview}) depicts this process: a rubric statement (e.g., “Describe the direction of dye movement”) maps to an SRG node (\textsc{Dye Direction}), which is then assigned a Bloom level (\textsc{Understand}). This visual link between task, rubric, SRG structure, and Bloom’s hierarchy enhances transparency for both learners and instructors.

\subsection{Sketch Reasoning Graphs (\texttt{SRG}s)}

We define an \texttt{SRG} as a cognitively annotated semantic graph extracted from a sketch:
\begin{equation}
f_{\mathrm{SRG}}(x) = G = (V, E, \ell, \lambda),
\end{equation}
where $ x \in \mathcal{I} $ is a sketch image, $ V \subseteq C $ are concepts from ontology $ \mathcal{O} = (C\text{oncept}, R\text{elation}) $, $ E \subseteq V \times V $ are directed relations such as causality, and $ \ell: V \to \mathcal{B} $ maps each node to a Bloom level. The annotation function $ \lambda: V \cup E \to \mathcal{E} $ captures visual and textual evidence ($\mathcal{E}$) supporting the cognitive label.

We treat both the rubric sketch $ r $ and student sketch $ x $ as inputs to this mapping, producing a reference graph $ G_o $ and a student graph $ G_s $ respectively. Importantly, the Bloom level annotations are central to how these graphs are constructed and interpreted throughout the system.

\subsection{Agent Roles}

\begin{figure}[htp!]
    \centering
    \begin{minipage}[b]{0.47\textwidth}
        \centering
        \begin{subfigure}[b]{\textwidth}
            \includegraphics[width=\textwidth]{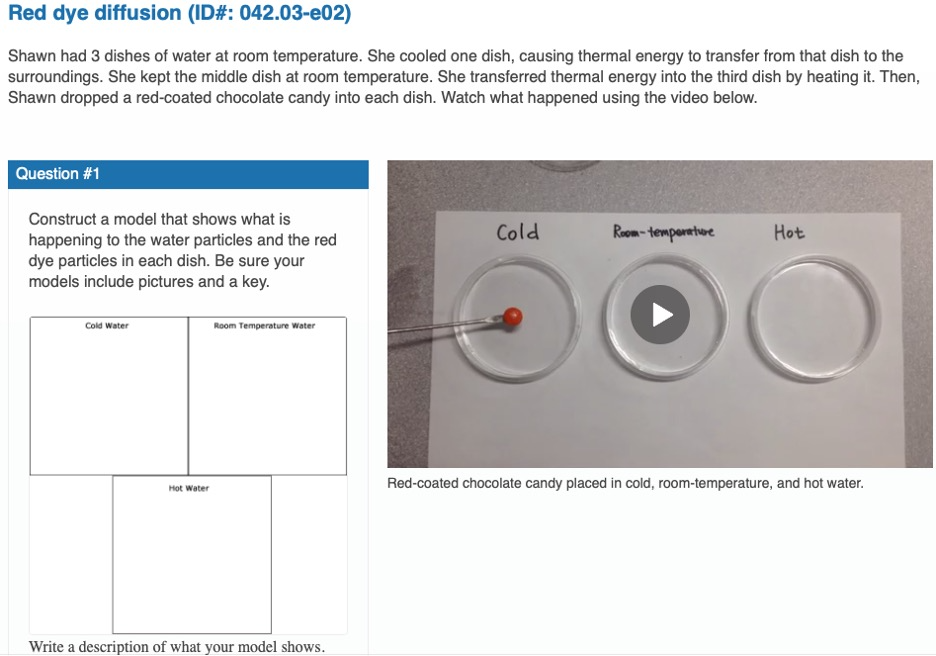}
            \caption{Question Snippet}
        \end{subfigure}
        \begin{subfigure}[b]{\textwidth}
            \includegraphics[width=\textwidth]{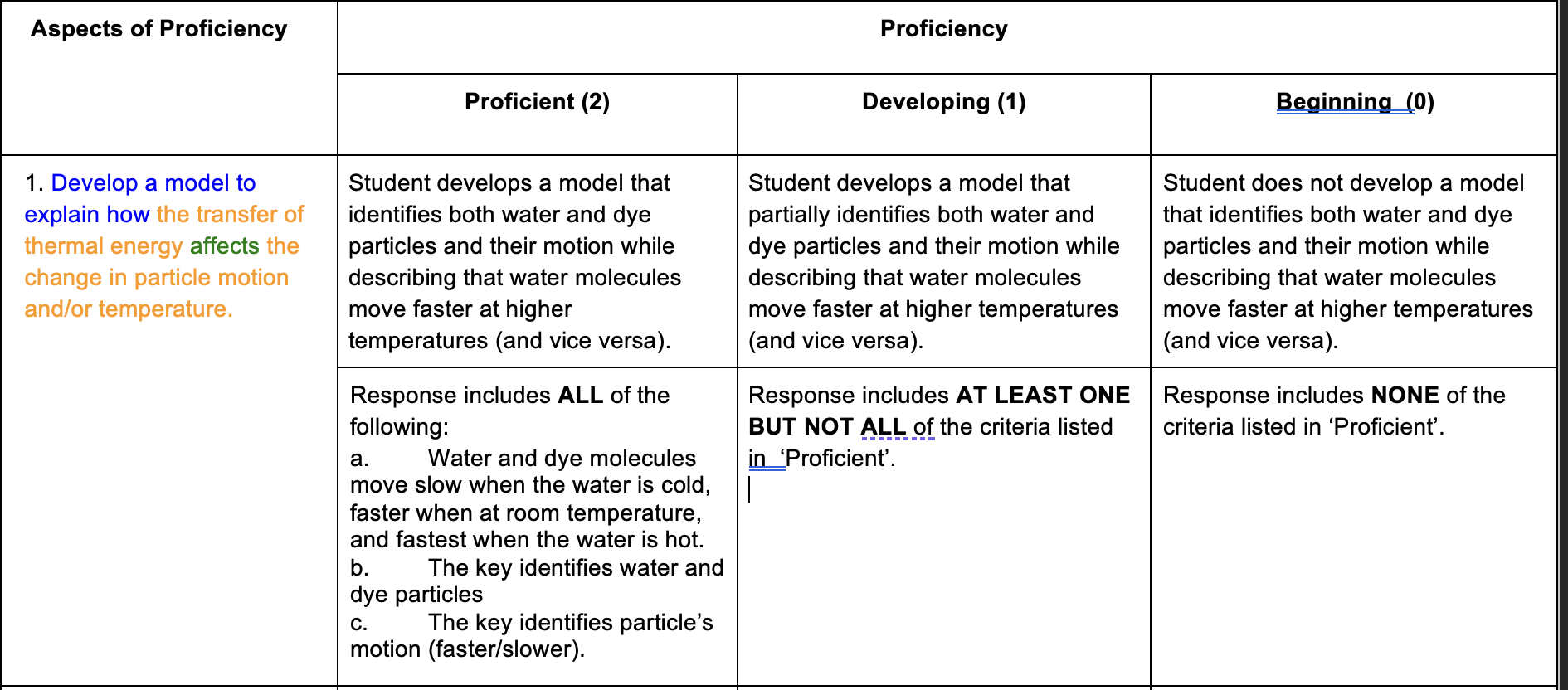}
            \caption{Raw Textual Rubric}
            \label{fig:SRG_overview:rubric}
        \end{subfigure}
    \end{minipage}
    \hfill
    \begin{minipage}[b]{0.52\textwidth}
        \centering
        \begin{subfigure}[b]{\textwidth}
            \includegraphics[width=\textwidth]{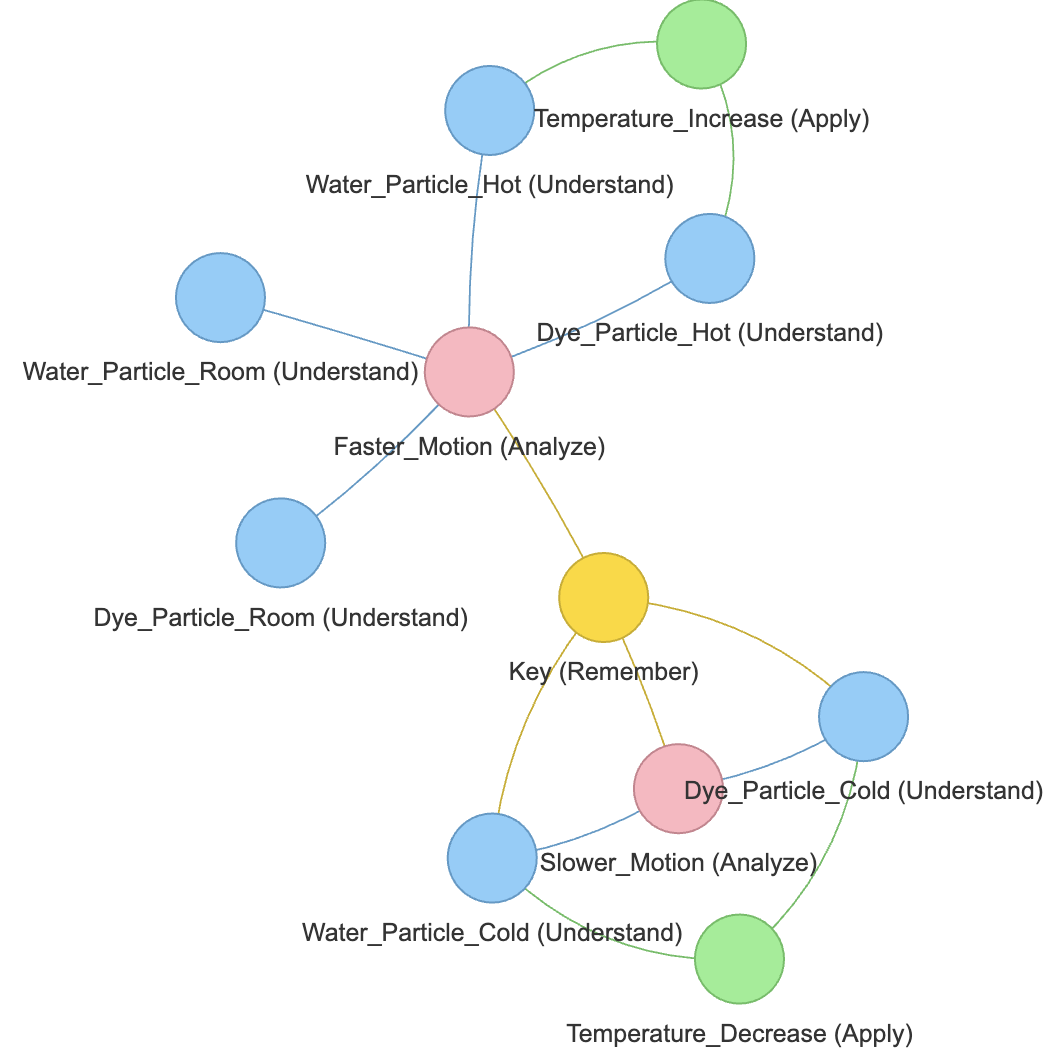}
            \caption{Agent1's Generated \texttt{SRG} (Gold-Standard)}
        \end{subfigure}
    \end{minipage}
    \caption{\textbf{Overview of \texttt{SRG} Generation.} Given a multi-model question including image, textual description of question and an expert-designed textual rubric for student sketch performance evaluation, Agent 1 processes the information and extracts \texttt{SRG} components and builds Level 4 Bloom's taxonomy ordered \texttt{SRG} (Bloom level ) to set the Gold standard for further evaluation and sketch modification.}
    \label{fig:SRG_overview}
    \vspace{-0.5cm}
\end{figure}

\textsc{SketchMind} comprises four agents, each of which contributes to the construction, interpretation, or refinement of \texttt{SRG}s for a Bloom-aligned assessment practice:

\paragraph{Agent 1: Rubric Parser.}
Agent 1 in \textsc{SketchMind} performs a static analysis of the rubric sketch $ r $ to construct $ G_o $, the gold-standard \texttt{SRG}. This process includes explicit mapping of each rubric concept to a Bloom level using expert annotations. For example, components that recall facts are labeled \textsc{Remember}, while those requiring functional understanding or multi-step reasoning are labeled \textsc{Apply} or \textsc{Analyze}. This Bloom-informed rubric becomes the benchmark for evaluating conceptual depth. Figure~\ref{fig:SRG_overview} delineates the \texttt{SRG} creation from the given question and textual rubric. Agent 1 not only generates gold-standard $G_o$ but also provides reverse mappings $\phi$ to create visuals from cognitive concepts for the target question. These reverse mappings will be used by subsequent agents to provide visual support and sketch modification for improved learning.

\begin{figure}[htp!]
    \centering
    \begin{subfigure}[b]{0.35\textwidth}
        \includegraphics[width=\linewidth]{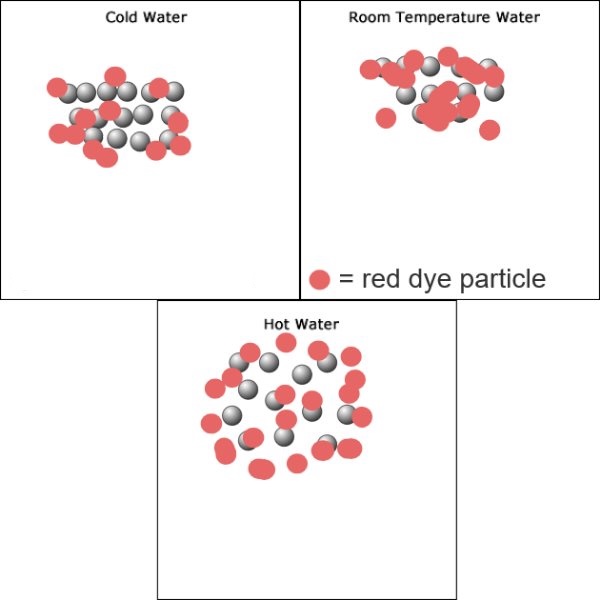}
        \caption{Input to Agent 2 (Student's drawn sketch)}
    \end{subfigure}
    \hfill
    \begin{subfigure}[b]{0.55\textwidth}
        \includegraphics[width=\linewidth]{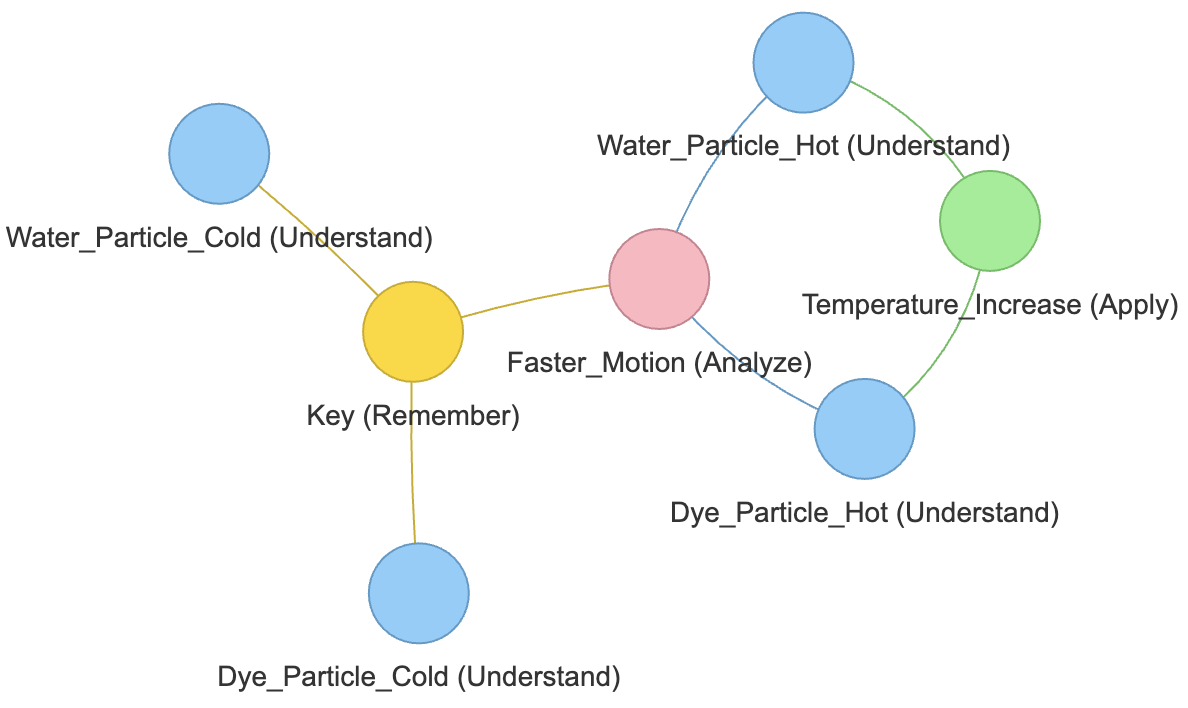}
        \caption{Agent 2's Perceived \texttt{SRG}}
    \end{subfigure}

    \caption{Sample sketch drawn by student and Agent 2 to extract perceived \texttt{SRG}. (a) student's drawn sketch, (b) Agent 2's perceived \texttt{SRG} based on the given sketch.}
    \vspace{-0.5cm}
    \label{fig:agent2_process}
\end{figure}

\paragraph{Agent 2: Perception.}
Agent 2 in \textsc{SketchMind} applies a vision-language model to infer the student \texttt{SRG} $ G_s $ from the sketch image $ x $. Beyond identifying visual elements, it infers semantic roles and Bloom levels using a classifier trained on labeled examples. For instance, correctly labeling a diagram element might reflect \textsc{Understand}, whereas indicating a dynamic interaction (e.g., force, flow) might reflect \textsc{Apply} or higher. Thus, Agent 2 directly constructs the cognitive structure of the student’s mental model. Figure~\ref{fig:agent2_process} delineates the sample student drawn sketch and Agent 2's perceived \texttt{SRG}, which is then further used for cognitive alignment by Agent 3.

\paragraph{Agent 3: Cognitive Alignment Evaluator.}
Agent 3 in \textsc{SketchMind} compares $ G_s $ to $ G_o $, computing structural and semantic similarity while analyzing Bloom-level mismatches. To get the similarity score, it first computes the ontology-based node alignment in such a way that for each pair $(v_s, v_o) \in V_s \times V_o$, it calculates weight $w(v_s, v_o)$ as:
\begin{equation}
w(v_s, v_o) = \alpha \cdot \text{sim}_\mathcal{O}(v_s, v_o) + (1 - \alpha) \cdot \mathbb{I}[\ell(v_s) = \ell(v_o)],
\end{equation}

where $\text{sim}_\mathcal{O}$ is semantic similarity from ontology $\mathcal{O}$. The summation of these weights is then normalized by the number of total nodes in both graphs to get the overall semantic similarity:
\begin{equation}
f_{OA}(V_s, V_o) = \frac{1}{|V_s + V_o|} \sum_{(v_s, v_o) \in \text{align}} w(v_s, v_o).
\end{equation}

The similarity score $ S \in [0,1] $ is defined as:
\begin{equation}
S(G_s, G_o) = 1 - \left( \gamma_1 \cdot \frac{f_{GED}(G_s, G_o)}{Z} + \gamma_2 \cdot (1 - f_{OA}(V_s, V_o)) \right),
\end{equation}
where $ f_{GED} $ is function to calculate graph-edit distance (node/edge insertions, deletions, substitutions) noramlized by $Z$ (total number of edges and nodes in both graphs), $ f_{OA} $ measures semantic alignment via ontology-based node similarity, and weights $ \gamma_1, \gamma_2 $ are calibrated on a training set. Notably, both the edit and alignment components consider Bloom-level mismatches as part of the error cost, penalizing regressions in cognitive complexity. It then computes the dominant Bloom level expressed in the sketch as:
\begin{equation}
\hat{y} = \mathrm{mode}\{\ell(v) \mid v \in V_s \cap V_o \iff S(G_s,G_o) > \tau \},
\end{equation}
where $\tau$ is the minimum similarity threshold pre-defined to extract overlapping features. This highlights any regression in complexity (e.g., if a concept expected at \textsc{Analyze} is represented at only \textsc{Remember}). These mismatches guide the diagnosis of underdeveloped concepts, forming the basis for targeted, Bloom-aligned feedback.

\begin{figure}[H]
    \centering
        \centering
        \begin{tcolorbox}[colback=white, colframe=black, title={Textual Feedback from Agent 3}]
        \small
        \textbf{Similarity\_score:} 0.592

        \textbf{Feedback:}

        \textbf{Your Proficiency Level:} Developing

        \textbf{What You Did Well:}\\
        The student demonstrated sound structure despite missing some components.

        \textbf{What Needs Attention:}
        \begin{itemize}
            \item Missing Concepts: Dye\_Particle\_Room, Temperature\_Decrease, and Slower\_Motion
        \end{itemize}

        \textbf{modification Guidance (Next Sketch Revisions by Visual Hint):}
        \begin{itemize}
            \item Water Particle Room (understand): Add markup to highlight water particles/molecules on the first block.
        \end{itemize}

        \textbf{Reasoning Gaps Detected In:}
        \begin{itemize}
         \item Dye\_Particle\_Room, Temperature\_Decrease, and Slower\_Motion
        \end{itemize}
        \end{tcolorbox}

    \caption{Cognitive alignment score and feedback for the perceived \texttt{SRG} (See Figure~\ref{fig:agent2_process}) generated by Agent 3 after similarity score calculations.}
    \label{fig:agent3_score_feedback}
\end{figure}

\paragraph{Agent 4: Feedback Generator and Sketch Modification.}
Agent 4 in \textsc{SketchMind} initiates a feedback loop when the similarity score $ S(G_s, G_o) $ falls below a threshold $ \tau $ (pre-defined to determine skecth proficiency). In our case, the curriculum expert defined three levels of proficiencies (Beginning, Developing, and Proficient) as can be seen in Figure~\ref{fig:SRG_overview:rubric}. $ \tau $ value is carefully calculated for each level and used by the agent. The agent identifies missing or misaligned nodes and edges, and traces each to its expected Bloom level. Using a trained reverse mapping $ \phi \colon (v, e) \to \text{VisualHint} $, Agent 4 generates cognitively aligned suggestions. For instance, if a student omits a causal interaction labeled \textsc{Analyze}, the system may overlay an arrow with a textual prompt like “What causes this effect?”. Below is the step-wise procedure of the sketch revision loop:
Given $G_o$ and $x^{(0)}$:
\begin{enumerate}
  \item Compute $G_s^{(t)} = f_{\mathrm{SRG}}(x^{(t)})$.
  \item If $S(G_s^{(t)}, G_o) \geq \tau$, exit loop.
  \item Identify deficient elements:
  \[
  \Delta^{(t)} = \left\{v \in V_s \mid v \not\in \text{aligned nodes} \right\} \cup \left\{e \in E_s \mid e \not\in E_o^{(t)} \right\}.
  \]
  \item Use reverse mapping $\phi: (v, e) \rightarrow \text{visual hint}$ generated by Agent 1 to provide visual suggestions $H^{(t)}$.
  \item Render $H^{(t)}$ on sketch canvas and generate Python code to modify the canvas and run locally.
  \item Modify the canvas with updated overlay and prompting student to revise $x^{(t)} \rightarrow x^{(t+1)}$.
  \item Repeat until $S \geq \tau$ or maximum iterations $T_{\text{max}}$ reached.
\end{enumerate}

\begin{figure}[htp!]
    \centering
    \begin{subfigure}[b]{0.55\textwidth}
        \includegraphics[width=\linewidth]{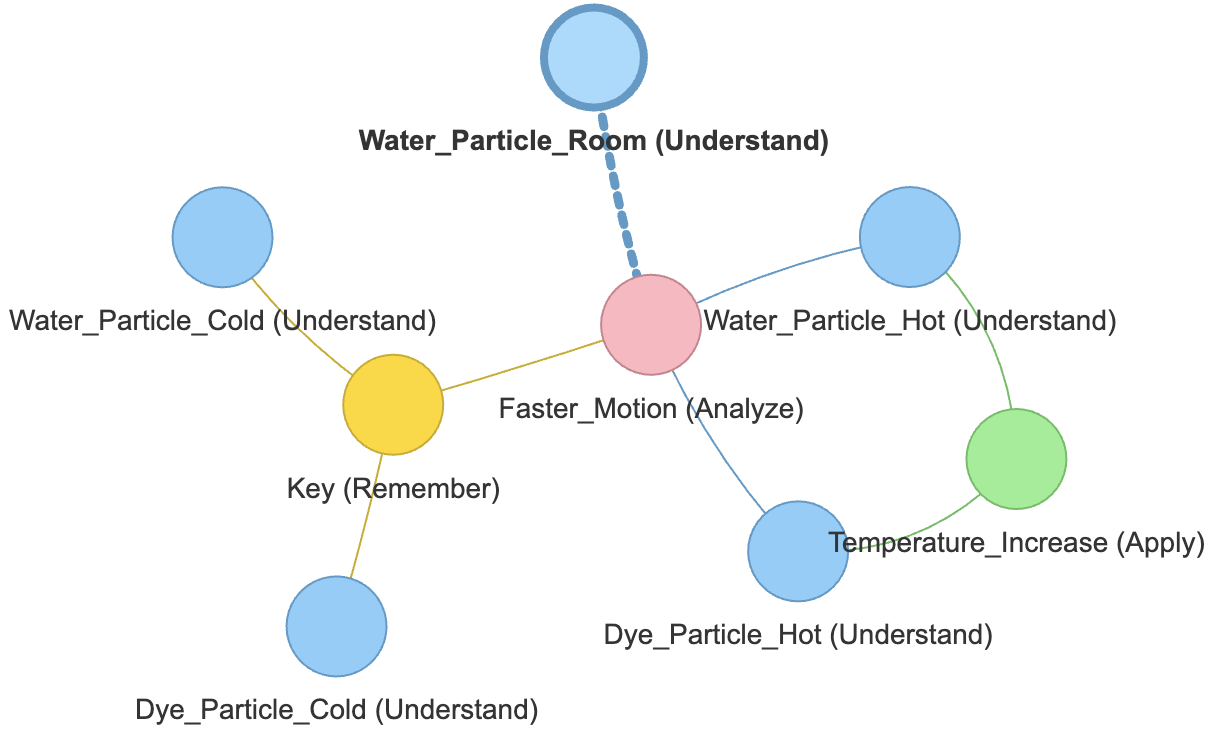}
        \caption{Modified \texttt{SRG} with Feedback from Agent 4}
    \end{subfigure}
    \hfill
    \begin{subfigure}[b]{0.35\textwidth}
        \includegraphics[width=\linewidth]{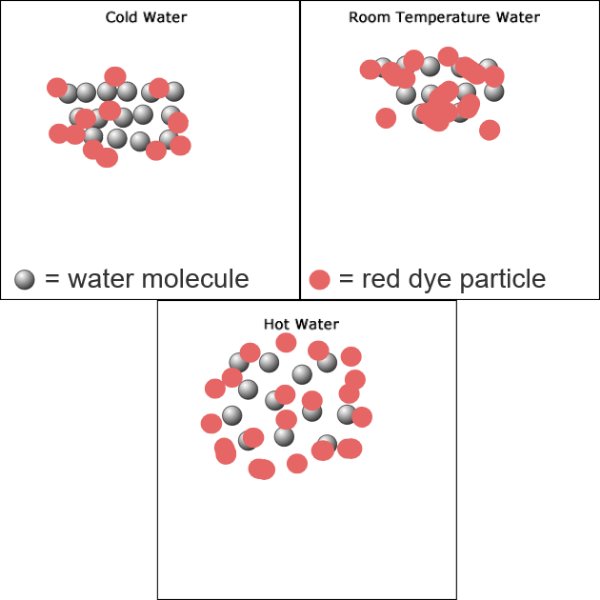}
        \caption{Agent 4's updated sketch for perceived feedback}
    \end{subfigure}

    \caption{Sample modified \texttt{SRG} based on the given feedback and score by Agent 3 and updated sketch with embedded python toolkit.}
    \label{fig:agent4_response}
\end{figure}

This sketch revision loop directly scaffolds the student toward higher-order cognitive tasks, iterating until the revised sketch meets conceptual fidelity (Modified \texttt{SRG} with additional node and updated sketch using the Python toolkit for the sample image given in Fig.~\ref{fig:agent4_response}). \textsc{SketchMind} is designed to adaptively scaffold students across diverse ability levels. If a sketch yields an incoherent or low-complexity \texttt{SRG} (e.g., all nodes at \textsc{Remember}), Agent 4 shifts focus to guided reconstruction, layering in increasingly complex prompts. The evaluator distinguishes perceptual errors from conceptual ones by analyzing visual evidence $ \lambda $, ensuring the feedback remains diagnostic rather than punitive. By explicitly modeling Bloom’s cognitive hierarchy at each stage of analysis—from rubric parsing to feedback generation \textsc{SketchMind} transforms sketch-based assessment into a learning-oriented, interpretable, and cognitively aligned process. \textsc{SketchMind} positions sketching not just as a representational task but as an active, assessable pathway for scientific reasoning and growth.

\section{Experimental Setup}
\label{sec:experimentalsetup}

We evaluate \textsc{SketchMind} using two distinct model configurations to comprehensively examine its performance and versatility across closed-source and open-source MLLM frameworks. In Configuration 1, we utilize GPT-4o, GPT-4.1, GPT-4.1-nano, o3, and o4-mini (all latest GPT models with different reasoning capabilities for comprehensive evaluation) accessed via the OpenAI API. This setup leverages black-box prompting tailored explicitly to encourage structured JSON output and support multimodal reasoning. Configuration 2 deploys open-source MLLMs, specifically the INT4-quantized Llama-4 Maverick and INT8-quantized Scout (400B and 109B parameters respectively, with 17B active parameters), running locally on four NVIDIA H100 GPUs, facilitated by Hugging Face Transformers (version 4.39+) and PyTorch (version 2.1+). We keep the same model for each agent to maintain the model performance consistency, for example, if targeted GPT-4o, then each agent uses GPT-4o and mainatin local chat session.

For closed-source configurations, inference requests are systematically routed through the OpenAI API, meticulously logging timestamps, model versions, and full prompt-response interactions to ensure transparency and reproducibility. Open-source models are executed on-premises, utilizing four H100 GPUs with CUDA 12.1 and cuDNN 8.9, loading quantized model weights from the Hugging Face Model Hub, and multimodal data inputs are managed by the \texttt{Llama4Processor}.

Both configurations operate with a multi-agent pipeline detailed in Section~\ref{sec:proposed_approach}, where all the agents run in sequence. Agents 1, 2, and 4 incorporate specifically designed prompt templates optimized for effective instruction-following, structured output enforcement, and multimodal reasoning integration. Agent 3 functions as a deterministic and model-agnostic component, focusing exclusively on graph comparison tasks. Comprehensive documentation of prompt templates is available in supplementary materials to enable precise experimental replication.

\paragraph{Dataset and Evaluation.}
\label{para:dataset_evaluation}

Despite growing interest in applying machine learning to educational domains, there remains a significant lack of publicly available, high-quality datasets that capture student-generated visual reasoning, particularly scientific sketches. To our knowledge, no large-scale, general-purpose benchmark exists that includes both raw student-drawn diagrams and structured expert annotations for meaningful semantic evaluation.

Given this gap, we base our study on a rigorously developed dataset originally introduced by Zhai et al.~\cite{zhai2022applying}, which has since become one of the most widely recognized resources for evaluating automated reasoning over student-generated scientific models. This dataset, adapted from the NGSA (Next Generation Science Assessment) initiative \cite{harris2024creating}, aligns closely with the NGSS framework \cite{ngss2013next} and has been used by AIED researchers to assess students’ conceptual understanding through multimodal evidence \cite{lee2023gemini,lee2023nerif}. Comprehensive details about dataset selection rationale and statistics are provided in Appendix~\ref{sec:appendix:datase_stats}. To ensure pedagogical validity, we conducted a structured human evaluation of model-generated feedback. Four domain-expert educators, each with graduate-level training in science education, independently assessed the pedagogical quality of system responses. Each rater evaluated a stratified random sample of 890 student-generated sketches (25\% of the dataset), ensuring balanced representation across baseline models (GPT-4o, GPT-4.1, O3), grade levels, and science task types. 

All evaluators participated in a calibration phase: they jointly annotated and discussed 10 representative examples to align rubric interpretation, and then independently scored the same 10 sketches. Inter-rater reliability, measured using Quadratic Weighted Kappa, reached $\kappa=0.83$, which is consistent with benchmarks for expert judgment in science assessment \cite{hickling2013applicability}. For the main study, each response was rated by at least two experts in a double-blind manner.

Evaluation guidelines were based on an extended rubric adapted from Zhai et al.~\cite{zhai2022applying} and tailored for formative science assessment. Experts rated each response along three dimensions: \textbf{1) Clarity:} Whether the textual and visual feedback was understandable and actionable. \textbf{2) Conceptual Accuracy:} Whether suggested modifications and explanations correctly reflected the target scientific concepts. And \textbf{3)Instructional Value:} The feedback’s potential to promote student learning and progression along Bloom’s taxonomy.  Each dimension was scored on a 5-point ordinal scale (1–5). Consistency across raters was supported through the initial calibration process, and ratings were conducted under double-blind conditions to mitigate bias. Framework–human agreement rates, along with detailed rubric codebooks and annotated examples, are reported in Section~4 and the supplementary material.

\paragraph{Implementation Details.}
The \texttt{SRG} construction pipeline in \textsc{SketchMind} utilizes a shared \texttt{SRGBuilder} class, which efficiently constructs, validates, and caches graphs, significantly reducing computational overhead and cost during repeated evaluations. Sketch adequacy is determined by a similarity threshold ($\tau=0.75$), with dynamic generation of visual hints guided by a reverse mapping ($\phi$) embedded within Agent 1’s implementation. We calculate the sketch prediction accuracy for each assessment item by comparing with human-expert annotated proficiencies as ($\text{Sum of correctly predicted samples across each proficiency level} / \text{Total samples}$) and average for all items as ($\text{Sum of all item's accuracies}/\text{Total number of items}$). We have evaluated the performance of \textsc{SkecthMind} by decomposing it into combination of target model with proposed \texttt{SRG}. This decomposition can help us understand the impact of target model and proposed \texttt{SRG} to determination best possible combination for \textsc{SketchMind}. Detailed evaluation scripts and additional implementation specifics are provided comprehensively in Appendix A.

\section{Results}
\label{sec:results}

Table~\ref{tab:accuracy_comparison} presents item-wise and macro-average accuracy across a range of MLLMs, both with and without \texttt{SRG} integration. The results consistently demonstrate that incorporating \texttt{SRG} supervision significantly improves performance across all models and items. For instance, GPT-4o, shows a substantial accuracy increase from 47.7\% to 76.5\% on Item H4-1, a relative gain of nearly 30 percentage points. Averaged across all items, GPT-4o benefits from an improvement of approximately 21.4\%. Even state-of-the-art models such as GPT-4.1 show meaningful accuracy gains ranging from 11.0\% to 15.4\% when \texttt{SRG} guidance is applied, with performance increasing from 74.2\% to 89.6\% on Item R1-1 and from 73.1\% to 87.2\% on Item H4-1. Hence, \textsc{SketchMind} works best with GPT-4.1 integrated with \texttt{SRG}.

Models with lower baseline performance, such as LLaMA 4 Maverick and Scout, experience even greater relative improvements. LLaMA 4 Maverick, for example, improves by 29.7\% on Item J2-1 and achieves up to 26.0\% gains on Item H4-1, suggesting that structured supervision via \texttt{SRG}s can dramatically enhance reasoning capabilities in non-reasoning models for open-source \textsc{SketchMind}.

\begin{table}[ht]
\centering
\caption{Item-wise accuracy (\%) across models with and without \texttt{SRG} integration for \textsc{SketchMind}.}
\label{tab:accuracy_comparison}
\begin{tabular}{l|c c c c c c | c}
\toprule
\textbf{Model} & \textbf{R1-1} & \textbf{J2-1} & \textbf{M3-1} & \textbf{H4-1} & \textbf{H5-1} & \textbf{J6-1} & \textbf{Average}\\
\midrule
GPT-4o & 63.2 & 58.4 & 53.5 & 47.7 & 52.3 & 58.6 & 55.6 \\
+ \texttt{SRG} & \textbf{78.5} & \textbf{77.4} & \textbf{75.8} & \textbf{76.5} & \textbf{74.7} & \textbf{79.1} & \textbf{77.1} \\
\textit{Gain} & \textit{\textcolor{Green}{+15.3}} & \textit{\textcolor{Green}{+19.0}} & \textit{\textcolor{Green}{+22.3}} & \textit{\textcolor{Green}{+28.8}} & \textit{\textcolor{Green}{+22.4}} & \textit{\textcolor{Green}{+20.5}} & \textit{\textcolor{Green}{+21.4}} \\
\midrule
GPT-4.1 & 74.2 & 78.5 & 77.4 & 73.1 & 79.6 & 81.5 & 77.4 \\
+ \texttt{SRG} & \textbf{\underline{89.6}} & \textbf{\underline{91.6}} & \textbf{\underline{88.4}} & \textbf{\underline{87.2}} & \textbf{\underline{91.7}} & \textbf{\underline{92.6}} & \textbf{\underline{90.2}} \\
\textit{Gain} & \textit{\textcolor{Green}{+15.4}} & \textit{\textcolor{Green}{+13.1}} & \textit{\textcolor{Green}{+11.0}} & \textit{\textcolor{Green}{+14.1}} & \textit{\textcolor{Green}{+12.1}} & \textit{\textcolor{Green}{+11.1}} & \textit{\textcolor{Green}{+12.8}} \\
\midrule
GPT-4.1-nano & 62.5 & 61.3 & 59.6 & 57.3 & 63.8 & 67.5 & 62.0 \\
+ \texttt{SRG} & \textbf{73.7} & \textbf{72.6} & \textbf{70.4} & \textbf{68.7} & \textbf{78.5} & \textbf{79.3} & \textbf{73.9} \\
\textit{Gain} & \textit{\textcolor{Green}{+11.2}} & \textit{\textcolor{Green}{+11.3}} & \textit{\textcolor{Green}{+10.8}} & \textit{\textcolor{Green}{+11.4}} & \textit{\textcolor{Green}{+14.7}} & \textit{\textcolor{Green}{+11.8}}  & \textit{\textcolor{Green}{+11.9}} \\
\midrule
O3 & 75.2 & 79.5 & 76.4 & 75.3 & 77.5 & 79.4 & 77.2\\
+ \texttt{SRG} & \textbf{89.5} & \textbf{91.1} & \textbf{87.3} & \textbf{86.4} & \textbf{89.6} & \textbf{90.3} & 89.0\\
\textit{Gain} & \textit{\textcolor{Green}{+14.3}} & \textit{\textcolor{Green}{+11.6}} & \textit{\textcolor{Green}{+10.9}} & \textit{\textcolor{Green}{+11.1}} & \textit{\textcolor{Green}{+12.1}} & \textit{\textcolor{Green}{+10.9}} & \textit{\textcolor{Green}{+11.8}} \\
\midrule
O4-mini & 71.4 & 75.3 & 73.3 & 69.2 & 74.6 & 76.1 & 73.3\\
+ \texttt{SRG} & \textbf{79.5} & \textbf{81.5} & \textbf{79.4} & \textbf{77.6} & \textbf{82.8} & \textbf{83.9} & \textbf{80.8} \\
\textit{Gain} & \textit{\textcolor{Green}{+8.1}} & \textit{\textcolor{Green}{+6.2}} & \textit{\textcolor{Green}{+6.1}} & \textit{\textcolor{Green}{+8.4}} & \textit{\textcolor{Green}{+8.2}} & \textit{\textcolor{Green}{+7.8}} & \textit{\textcolor{Green}{+7.5}} \\
\midrule
LLaMA 4 Scout & 48.6 & 43.2 & 39.5 & 38.4 & 45.6 & 47.5 & 43.8\\
+ \texttt{SRG} & \textbf{63.8} & \textbf{69.3} & \textbf{59.6} & \textbf{61.7} & \textbf{67.5} & \textbf{66.8} & \textbf{64.8} \\
\textit{Gain} & \textit{\textcolor{Green}{+15.2}} & \textit{\textcolor{Green}{+26.1}} & \textit{\textcolor{Green}{+20.1}} & \textit{\textcolor{Green}{+23.3}} & \textit{\textcolor{Green}{+21.9}} & \textit{\textcolor{Green}{+19.3}}  & \textit{\textcolor{Green}{+21.0}} \\
\midrule
LLaMA 4 Maverick & 53.4 & 49.7 & 44.5 & 42.7 & 46.8 & 49.6 & 47.8\\
+ \texttt{SRG} & \textbf{77.3} & \textbf{79.4} & \textbf{63.6} & \textbf{68.7} & \textbf{71.8} & \textbf{73.5} & \textbf{72.4}\\
\textit{Gain} & \textit{\textcolor{Green}{+23.9}} & \textit{\textcolor{Green}{+29.7}} & \textit{\textcolor{Green}{+19.1}} & \textit{\textcolor{Green}{+26.0}} & \textit{\textcolor{Green}{+25.0}} & \textit{\textcolor{Green}{+23.9}} & \textit{\textcolor{Green}{+24.9}}\\
\bottomrule
\end{tabular}
\end{table}

\paragraph{Why Multi-Agent Framework?} All the above-mentioned results are performed with the multi-agent framework proposed for \textsc{SketchMind}, but here comes the question: why not a single agent? To answer that and assess the impact of modularization inherent in the mult-agent framework for reasoning tasks, we conducted an ablation study comparing a unified single-agent framework with our proposed multi-agent pipeline. Each setting was evaluated with and without \texttt{SRG} supervision using two strong backbone models: GPT-4o and GPT-4.1.

The results in Table~\ref{tab:ablation_results} demonstrate that modularizing the reasoning process via a multi-agent framework consistently improves \textsc{SketchMind}'s performance over the single-agent baseline across all six items. Without \texttt{SRG} integration, GPT-4o’s accuracy increases from 50.1\% (single-agent) to 55.6\% (multi-agent), while GPT-4.1 improves from 62.9\% to 77.4\%, indicating that decomposing tasks into specialized agents enables more structured, context-aware reasoning even without explicit graph guidance. This performance gap widens with \texttt{SRG} supervision: GPT-4o’s accuracy rises from 69.5\% to 77.1\% and GPT-4.1 from 82.8\% to 90.2\%, with item-wise gains ranging from 5.2\% to 13.4\%. Notably, Item H4-1 (Hot Shower Effect), which demands the highest Bloom’s taxonomy level (Create), sees accuracy climb from 63.2\% to 76.5\% for GPT-4o and from 79.3\% to 87.2\% for GPT-4.1 when switching to multi-agent reasoning with SRG, confirming that modular agents are better equipped to leverage structured semantic guidance. These results highlight that a multi-agent architecture, where reasoning responsibilities are explicitly segmented and coordinated, facilitates more robust and interpretable scientific reasoning.

\begin{table}[htp!]
\centering
\caption{Ablation study comparing single-agent and multi-agent frameworks for \textsc{SketchMind} (accuracy in \%)}
\label{tab:ablation_results}
\begin{tabular}{l|c c c c c c | c}
\toprule
\textbf{Model configurations} & \textbf{R1-1} & \textbf{J2-1} & \textbf{M3-1} & \textbf{H4-1} & \textbf{H5-1} & \textbf{J6-1} & \textbf{Average} \\
\midrule
GPT-4o (Single Agent w/o SRG) & 56.3 & 52.4 & 49.3 & 43.1 & 48.3 & 51.2 & 50.1 \\
GPT-4o (Multi-Agent w/o SRG) & 63.2 & 58.4 & 53.5 & 47.7 & 52.3 & 58.6 & 55.6 \\
GPT-4o (Single Agent w/ SRG) & 73.5 & 69.6 & 68.4 & 63.2 & 68.8 & 74.3  & 69.5\\
GPT-4o (Multi-Agent w/ SRG) & \textbf{78.5} & \textbf{77.4} & \textbf{75.8} & \textbf{76.5} & \textbf{74.7} & \textbf{79.1} & \textbf{77.1} \\
\midrule
GPT-4.1 (Single Agent w/o SRG) & 69.6 & 61.3 & 59.2 & 57.1 & 58.8 & 71.2 & 62.9 \\
GPT-4.1 (Multi-Agent w/o SRG) & 74.2 & 78.5 & 77.4 & 73.1 & 79.6 & 81.5 & 77.4\\
GPT-4.1 (Single Agent w/ SRG) & 84.4 & 81.2 & 83.6 & 79.3 & 82.7 & 85.3 & 82.8 \\
GPT-4.1 (Multi-Agent w/ SRG) & \textbf{89.6} & \textbf{91.6} & \textbf{88.4} & \textbf{87.2} & \textbf{91.7} & \textbf{92.6} & \textbf{90.2} \\
\bottomrule
\end{tabular}
\end{table}
\paragraph{Feedback and Sketch Modification Evaluation.} To assess the pedagogical quality and usefulness of \textsc{SketchMind}'s generated sketches and feedback, we conducted a human evaluation study focusing on Agent 4, the component responsible for modification of visual representations and providing formative feedback. Expert evaluators, comprising experienced science educators, rated the outputs across six science items based on their clarity, conceptual accuracy, and instructional value.

As shown in Table~\ref{tab:feedback_eval}, GPT-4.1 achieved the highest average rating (4.1), followed closely by o3 (4.0) and LLaMA 4 Maverick (3.5). Evaluators consistently noted that sketches generated using GPT-4.1 were pedagogically sound, well-aligned with scientific principles, and accompanied by feedback that could directly support improved student learning. Lower-performing models, such as GPT-4o and LLaMA 4 Scout, received average ratings of 2.3 and 2.5, respectively, often due to missing or vague concepts and less actionable feedback. The evaluators emphasized that when integrated with high-performing language models like GPT-4.1, \textsc{SketchMind} has the potential to significantly improve the quality of student-generated scientific sketches through guided modification and tailored feedback. 

\begin{table}[ht]
\centering
\caption{Human Evaluation Ratings of Feedback and Sketch Modification (1 = Poor, 5 = Excellent).}
\label{tab:feedback_eval}
\begin{tabular}{l|c c c c c c|c}
\toprule
\textbf{Model} & \textbf{R1-1} & \textbf{J2-1} & \textbf{M3-1} & \textbf{H4-1} & \textbf{H5-1} & \textbf{J6-1} & \textbf{Average} \\
\midrule
GPT-4o & 2.5 & 2.0 & 2.5 & 2.5 & 2.5 & 2.0 & 2.3 \\
GPT-4.1 & \textbf{4.5} & \textbf{4.0} & \textbf{3.5} & \textbf{3.5} & \textbf{4.5} & \textbf{4.5} & \textbf{4.1} \\
GPT-4.1-nano & 3.0 & 3.0 & 3.5 & 2.5 & 3.0 & 3.5 & 3.1 \\
O3 & 4.5 & 4.0 & 4.0 & 4.0 & 3.5 & 4.0 & 4.0 \\
O4-mini & 4.0 & 3.5 & 3.0 & 3.0 & 3.5 & 3.0 & 3.3 \\
LLaMA 4 Scout & 3.0 & 2.5 & 2.0 & 2.5 & 2.0 & 3.0 & 2.5 \\
LLaMA 4 Maverick & 3.5 & 4.0 & 3.5 & 3.5 & 3.0 & 3.5 & 3.5 \\
\bottomrule
\end{tabular}
\end{table}

\section{Conclusion}

In this work, we introduced \textsc{SketchMind}, a cognitively grounded multi-agent system for assessing and improving student-generated scientific sketches. By leveraging \texttt{SRG}s annotated with Bloom’s taxonomy, \textsc{SketchMind} enables interpretable evaluation, pedagogically aligned feedback, and iterative modification of higher-quality visual explanations. Empirical results across six NGSS-aligned items show that \textsc{SketchMind} (model + \texttt{SRG}) substantially outperforms both MLLM baselines without \texttt{SRG} and single-agent pipelines, achieving up to 90.2\% average accuracy with GPT-4.1 and receiving high expert ratings of (4.1) for feedback quality. Overall, each MLLM shows significant improvement for sketch reasoning and prediction with the proosed \texttt{SRG} integration which highlights the significance of structural reasoning for scientific sketch evaluations. \textsc{SketchMind} bridges the gap between visual AI reasoning and education by embedding cognitive theories directly into the assessment pipeline.

\paragraph{Limitations and future directions.} \phantomsection
\label{para:limitations} Despite promising results, several limitations warrant attention. First, while the multi-agent system is modular, inter-agent coordination is static and predefined, future work could explore dynamic planning strategies using large language model (LLM) controllers or reinforcement learning, as demonstrated in recent advances in task decomposition for multi-agent collaboration \cite{geng2025realm, guo2024using}. Second, our findings relies on overal sketch prediction for given proficieicny level; however, \texttt{SRG}-level evaluation may provide in-depth analysis of predictions. Future work could involve experts to create manual \texttt{SRG}s for in-depth analysis. Lastly, Incorporating student behavioral data (e.g., stroke sequence or eye tracking) into the \texttt{SRG} modeling pipeline may also further enhance alignment with cognitive engagement signals \cite{kamalov2025evolution}.

\begin{ack}
The research reported here was supported by the Institute of Education Sciences, U.S. Department of Education, through Grant R305C240010 (PI Zhai). The opinions expressed are those of the authors and do not represent views of the Institute or the U.S. Department of Education. 
\end{ack}

\bibliographystyle{plainnat}
\bibliography{references}

\end{document}